\documentclass[12pt]{article}
\usepackage{amssymb,bm}
\usepackage{amsmath}
\usepackage{amsthm}
\usepackage{graphicx}
\usepackage[active]{srcltx} 
\usepackage{hyperref} 
\hypersetup{pdfborder=0 0 0}
\newtheorem{theorem}{{\sc Theorem}}[section]

\newcommand{\bb}[1]{\mathbb{ #1}}

\bmdefine\Bone{1}

\newcommand{\bra}[1]{\overline{#1}}

\newcommand{\nth}[1]{\displaystyle\frac{1}{#1}}

\renewcommand{\Hat}[1]{\widehat{#1}}

\def\XXint#1#2#3{{\setbox0=\hbox{$#1{#2#3}{\int}$ }
\vcenter{\hbox{$#2#3$ }}\kern-.6\wd0}}

\newcommand{\im}{\mathfrak{Im}}

\newcommand{\rhs}{right-hand side}

\newcommand{\Ga}{\alpha}
\newcommand{\Gb}{\beta}

\newcommand{\Ge}{\epsilon}
\newcommand{\eps}{\varepsilon}

\newcommand{\Gk}{\kappa}

\newcommand{\Gth}{\theta}

\newcommand{\Gs}{\sigma}

\newcommand{\Go}{\omega}

\newcommand{\GD}{\Delta}

\newcommand{\GO}{\Omega}

\bmdefine\BGa{\alpha}
\bmdefine\BGb{\beta}
\bmdefine\BGd{\delta}
\bmdefine\BGe{\epsilon}
\bmdefine\BGve{\varepsilon}
\bmdefine\BGf{\phi}
\bmdefine\BGvf{\varphi}
\bmdefine\BGg{\gamma}
\bmdefine\BGc{\chi}
\bmdefine\BGi{\iota}
\bmdefine\BGk{\kappa}
\bmdefine\BGl{\lambda}
\bmdefine\BGn{\eta}
\bmdefine\BGm{\mu}
\bmdefine\BGv{\nu}
\bmdefine\BGp{\pi}
\bmdefine\BGth{\theta}
\bmdefine\BGvth{\vartheta}
\bmdefine\BGr{\rho}
\bmdefine\BGvr{\varrho}
\bmdefine\BGs{\sigma}
\bmdefine\BGvs{\varsigma}
\bmdefine\BGt{\tau}
\bmdefine\BGj{\tau}
\bmdefine\BGu{\upsilon}
\bmdefine\BGo{\omega}
\bmdefine\BGx{\xi}
\bmdefine\BGy{\psi}
\bmdefine\BGz{\zeta}
\bmdefine\BGD{\Delta}
\bmdefine\BGF{\Phi}
\bmdefine\BGG{\Gamma}
\bmdefine\BGL{\Lambda}
\bmdefine\BGP{\Pi}
\bmdefine\BGT{\Theta}
\bmdefine\BGS{\Sigma}
\bmdefine\BGU{\Upsilon}
\bmdefine\BGO{\Omega}
\bmdefine\BGX{\Xi}
\bmdefine\BGY{\Psi}

\bmdefine\BFM{\mathfrak{M}}
\bmdefine\BFb{\mathfrak{b}}
\bmdefine\BFk{\mathfrak{k}}
\bmdefine\BFm{\mathfrak{m}}
\bmdefine\BFu{\mathfrak{u}}
\bmdefine\BFv{\mathfrak{v}}

\bmdefine\BCA{{\mathcal A}}
\bmdefine\BCB{{\mathcal B}}
\bmdefine\BCC{{\mathcal C}}
\bmdefine\BCD{{\mathcal D}}
\bmdefine\BCE{{\mathcal E}}
\bmdefine\BCF{{\mathcal F}}
\bmdefine\BCG{{\mathcal G}}
\bmdefine\BCH{{\mathcal H}}
\bmdefine\BCI{{\mathcal I}}
\bmdefine\BCJ{{\mathcal J}}
\bmdefine\BCK{{\mathcal K}}
\bmdefine\BCL{{\mathcal L}}
\bmdefine\BCM{{\mathcal M}}
\bmdefine\BCN{{\mathcal N}}
\bmdefine\BCO{{\mathcal O}}
\bmdefine\BCP{{\mathcal P}}
\bmdefine\BCQ{{\mathcal Q}}
\bmdefine\BCR{{\mathcal R}}
\bmdefine\BCS{{\mathcal S}}
\bmdefine\BCT{{\mathcal T}}
\bmdefine\BCU{{\mathcal U}}
\bmdefine\BCV{{\mathcal V}}
\bmdefine\BCW{{\mathcal W}}
\bmdefine\BCX{{\mathcal X}}
\bmdefine\BCY{{\mathcal Y}}
\bmdefine\BCZ{{\mathcal Z}}

\bmdefine\Bzr{ 0}
\bmdefine\Ba{ a}
\bmdefine\Bb{ b}
\bmdefine\Bc{ c}
\bmdefine\Bd{ d}
\bmdefine\Be{ e}
\bmdefine\Bf{ f}
\bmdefine\Bg{ g}
\bmdefine\Bh{ h}
\bmdefine\Bi{ i}
\bmdefine\Bj{ j}
\bmdefine\Bk{ k}
\bmdefine\Bl{ l}
\bmdefine\Bm{ m}
\bmdefine\Bn{ n}
\bmdefine\Bo{ o}
\bmdefine\Bp{ p}
\bmdefine\Bq{ q}
\bmdefine\Br{ r}
\bmdefine\Bs{ s}
\bmdefine\Bt{ t}
\bmdefine\Bu{ u}
\bmdefine\Bv{ v}
\bmdefine\Bw{ w}
\bmdefine\Bx{ x}
\bmdefine\By{ y}
\bmdefine\Bz{ z}
\bmdefine\BA{ A}
\bmdefine\BB{ B}
\bmdefine\BC{ C}
\bmdefine\BD{ D}
\bmdefine\BE{ E}
\bmdefine\BF{ F}
\bmdefine\BG{ G}
\bmdefine\BH{ H}
\bmdefine\BI{ I}
\bmdefine\BJ{ J}
\bmdefine\BK{ K}
\bmdefine\BL{ L}
\bmdefine\BM{ M}
\bmdefine\BN{ N}
\bmdefine\BO{ O}
\bmdefine\BP{ P}
\bmdefine\BQ{ Q}
\bmdefine\BR{ R}
\bmdefine\BS{ S}
\bmdefine\BT{ T}
\bmdefine\BU{ U}
\bmdefine\BV{ V}
\bmdefine\BW{ W}
\bmdefine\BX{ X}
\bmdefine\BY{ Y}
\bmdefine\BZ{ Z}

\title{A theory of inductive loops in electrochemical impedance spectroscopy}
\author{Yury Grabovsky\thanks{Department of Mathematics, Temple
    University, Philadelphia, PA 19122, USA.}\and Jacob
  Guynee\thanks{Department of Mathematics, Georgia Institute of Technology, Atlanta, GA 30332, USA.}}
\begin{document}
\maketitle
\begin{abstract}
  We demonstrate that failure of time-invariance assumption in the modeling of
  electrochemical systems by equivalent circuits can lead to the formation of
  low frequency ``inductive loops'' that manifest themselves as positive
  imaginary parts of the impedance function. Assuming that the properties of
  the equivalent circuits change slowly in time we perform an
  asymptotic analysis and obtain a new integral representation of
  the impedance function that reduces to the standard one at high
  frequencies, while exhibiting inductive loops at low frequencies.
\end{abstract}
\section{Introduction}
Electrochemical impedance spectroscopy (EIS) is an indispensable tool to
describe complex electrochemical systems in a unified and graphical way.
In this approach any system is described by a single complex-valued impedance
function $Z(\Go)$. The basic theory of EIS \cite{ssk93,bafa00,bamc05} says that this
function possesses special analytic properties that are most concisely
expressed by the representation  
\begin{equation}
  \label{specrep}
  Z(\Go)=\nth{iC_{0}\Go}+\int_{0}^{\infty}\frac{d\Gs(\tau)}{1+i\Go\tau},\qquad
\int_{0}^{\infty}\frac{d\Gs(\tau)}{1+\tau}<+\infty,\quad 0<C_{0}\le\infty.
\end{equation}
Here $\Gs$ is a positive Borel-regular measure on $[0,+\infty)$. It is often convenient
to  approximate such measures by linear combinations of delta-functions,
resulting in rational approximations of $Z(\Go)$ \cite{aogr92}. Such approximations can be
interpreted as impedances of electrical circuits made of resistors and
capacitors only. They are called the equivalent circuit models (ECM). One
easily verifiable feature of the representation (\ref{specrep}) is the
negativity of the imaginary part of the impedance function. 

The experimentally observed arcs in a Nyquist plot of the impedance with
positive imaginary part in the low frequency part of the spectrum received the
unfortunate moniker of an ``inductive loop''
\cite{gtida05,cps09,zqxqs12,tff16,crgr21}. It is generally agreed that
magnetic effects play no significant role in the functioning of most
electrochemical systems and devices, and even when they do, the inductive
effects manifest themselves at the high end of the frequency spectrum
\cite{bhw16,klotz19,bouk20}. The explanation of inductive loops in the current
literature \cite{klotz19,bouk20} by means of ECMs with negative resistances
and capacitances creates more problems than it solves. If one permits negative
resistances and capacitances, then \emph{arbitrary} ``experimental data'' can
be matched by such an impedance function with any degree of precision due to
the Riesz theorem (see e.g. \cite{part97,grho-CEMP,grho-gen}). This is not
surprising, since it is the passivity of the system
\cite{wobe65,cps02,sriv20}, violated by the the negative resistances and
capacitances in ECMs, that are responsible for the possibility of stable
reconstruction of the impedance function.

A far more reasonable explanation of the experimental appearance of
low frequency inductive loops is the violation of the time-invariance
assumption at time scales commensurate with the inverses of low
frequencies. This explanation is semi-explicit in
\cite{tff16,klotz19,shw20}, where specific electrochemical processes
altering the properties of the system on slow time scales are
identified.

When the time-invariance assumption is violated the input-output behavior of
the system can no longer be mathematically described by a well-defined
impedance function at low frequencies. In this paper we will show that if one
applies a particular impedance measurement procedure, delivering the
correct impedance for any time-invariant system, the inductive loops could be
observed in non time-invariant systems.

The goal of this paper is to extend the classical EIS/ECM approach to
electrochemical systems with slowly ``drifting'' properties. In
Section~\ref{sec:EVC} we demonstrate that a single Voigt element with drifting
properties may exhibit an inductive loop. In Section~\ref{sec:asym} we use
asymptotic analysis to extend the EIS/ECM approach to non time-invariant
systems with slowly varying properties. The resulting impedance representation
formula, though more complicated, may still be used to reconstruct the
impedance function values within the frequency band containing experimental
measurements by performing the classical Kramers-Kronig analysis
\cite{aogr95,bouk95,digr01,sriv20} for high frequency data, while fitting the
additional ``drift'' parameters using the inductive loop low frequency data. 

The paper is organized as follows. In Section~\ref{sec:meas} we
discuss a measurement technique that gives a good approximation of the
true impedance function for time-invariant systems. We then show in
Section~\ref{sec:EVC} that the same measurement technique for non
time-invariant Voigt elements can produce ``inductive loop''
data. Representing a general non time-invariant electrochemical system
by an ECM with non time-invariant Voigt elements connected in series,
gives us a mathematical model that can be analyzed. An asymptotic
analysis applied to each non time-invariant Voigt element leads us in
Section~\ref{sec:asym} to a generalization of the EIS theory that reduces to classical
formulas at high frequencies and is capable of modeling low frequency inductive loops.

\section{Impedance measurements}
\label{sec:meas}
Let us assume that the input signal is the current
\begin{equation}
  \label{Input}
I(t)=I_{0}\sin(\Go t),\quad t\in[0,T(\Go)],\quad T(\Go)=\frac{2\pi n(\Go)}{\Go},
\end{equation}
where $n(\Go)\in\bb{N}$ should be as large as possible, so that the time
$T(\Go)$ it takes to make a measurement is still acceptable to whoever makes
the measurements. Of course, this is an issue only for low frequencies
$\Go$. Thus, we can measure the system response only for frequencies
$\Go\ge\Go_{\min}=2\pi/T_{\max}$, where $T_{\max}$ is the maximal admissible
time for making low frequency measurements. Hence, in the low frequency regime
we always choose $n(\Go)=1$. At high frequencies we use a
different measurement strategy. We choose a fixed measurement time $T_{0}$ and
set $n(\Go)=$\texttt{round}$(T_{0}\Go/2\pi)$. Hence, for all $\Go\ge\Go_{\min}$
we define
\begin{equation}
  \label{nofw}
n(\Go)=\max\left\{1,\texttt{round}\left(\frac{T_{0}\Go}{2\pi}\right)\right\},\qquad
\Go\ge\Go_{\min}.
\end{equation}
The Fourier transform of $I(t)$ can be computed explicitly, but is
an unwieldy expression, except at the frequency $\Go$ of the input sinusoid:
\begin{equation}
  \label{FTinput}
  \Hat{I}(\Go)=I_{0}\int_{0}^{T(\Go)}\sin(\Go t)e^{-i\Go t}dt=
\frac{\pi I_{0}n}{i\Go}=\frac{I_{0}T(\Go)}{2i}.
\end{equation}
For linear, time-invariant (LTI) systems the output voltage $U(t)$ must
satisfy $\Hat{U}(\Go)=Z(\Go)\Hat{I}(\Go)$, where the complex
factor $Z(\Go)$ is called the impedance. Theoretically, when $I(t)=0$ for
$t>T(\Go)$, the output voltage is still non-zero. However, in most
cases it decays exponentially fast and can therefore be neglected. In this
case we have the approximation
\begin{equation}
  \label{FToutput}
  \Hat{U}(\Go)\approx\int_{0}^{T(\Go)}U(t)e^{-i\Go t}dt.
\end{equation}
Since $U(t)$ is measured, formula (\ref{FToutput}) can be used to estimate
$\Hat{U}(\Go)$ and thus, we have the formula for estimating the impedance
from the experimental data:
\begin{equation}
  \label{impform}
Z_{0}(\Go)=\frac{2i}{T(\Go)}\int_{0}^{T(\Go)}U(t)e^{-i\Go t}dt.
\end{equation}
\begin{figure}[t] 
  \centering
  \includegraphics[scale=0.35]{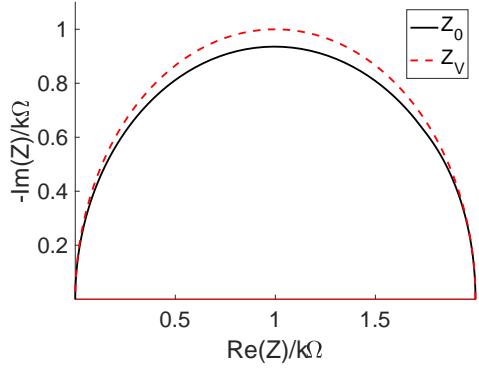}
  \caption{Comparison of $Z_{0}(\Go)$ and $Z(\Go)$ for an elementary Voigt
    circuit with $R=2k\GO$, $C=200\mu F$.}
  \label{fig:Z0vsZV}
\end{figure}
If we apply formula (\ref{impform}) to an elementary
Voigt circuit, consisting of a resistor $R$ and a capacitor $C$ connected in
parallel, we \emph{will not} get the correct answer
\begin{equation}
  \label{ZV}
  Z_{V}(\Go)=\frac{R}{i\Gth\Go+1},\qquad\Gth=RC.
\end{equation}
Figure~\ref{fig:Z0vsZV} shows that  $Z_{0}$ is not a very good approximation
for $Z_{V}$, especially in the intermediate frequency range.
The remedy is to understand this discrepancy and then
devise a way to correct for it. Using the explicit
expression for $Z_{0}(\Go)$ (which we don't display here), we
discover that for an elementary Voigt circuit we have
\begin{equation}
  \label{Z0Zrel}
    Z_{0}(\Go)=Z_{V}(\Go)-\left(\frac{1}{T\Go}\im(Z_{V}(\Go))+\frac{i}{T}Z'_{V}(\Go)\right)
\left(1-e^{-\frac{T}{\Gth}}\right),
\end{equation}
where $Z'_{V}(\Go)$ denotes the derivative of $Z_{V}(\Go)$ with respect to
$\Go$. Since relation (\ref{Z0Zrel}) between the measured and the true
impedance function of an elementary Voigt circuit is linear, it extends to
\emph{all LTI systems}. 

We recall that in an LTI system the output (voltage
$U(t)$) depends on the input (current $I(t)$) via
\begin{equation}
  \label{LTI}
  U(t)=\rho_{0}I(t)+\int_{-\infty}^{t}I(\tau)K(t-\tau)d\tau,
\end{equation}
where the function $K(s)$ is called a memory kernel.
\begin{theorem}
\label{th:Zest}
  Suppose that the memory kernel decays exponentially:
  \begin{equation}
    \label{decay}
    |K(s)|\le\frac{R_{0}}{\Gth}e^{-s/\Gth},\qquad s>T_{0}.
  \end{equation}
for some $T_{0}<T(\Go)$ for all $\Go$. Then
\begin{equation}
  \label{correst}
\left|Z_{0}(\Go)-Z(\Go)+\frac{1}{T\Go}\im(Z(\Go))+\frac{i}{T}Z'(\Go)\right|\le 
R_{0}\left(1+\frac{\Gth}{T}\right)e^{-T/\Gth},
\end{equation}
where
\begin{equation}
  \label{trueZ}
  Z(\Go)=\rho_{0}+\int_{0}^{\infty}K(s)e^{-i\Go s}ds
\end{equation}
is the true impedance of the system.
\end{theorem}
The proof can be found in Appendix~\ref{app:proofs}

Thus, we are lead to a simple method for estimating the
true impedance by making measurements over a longer time
$T_{2}=2\pi n_{2}(\Go)/\Go>T_{1}=2\pi n_{1}(\Go)/\Go$. Then, up to an
exponentially small error we must have
\begin{equation}
  \label{Zcorr}
  Z(\Go)\approx Z_{\rm exp}(\Go)=\frac{T_{2}Z_{0}(\Go;T_{2})-T_{1}Z_{0}(\Go;T_{1})}{T_{2}-T_{1}}.
\end{equation}
We still want to require that at high frequencies the time it takes to make a
an impedance measurement is fixed. Let us call it $T_{0}'$, since we still want
$n_{1}(\Go)$ to be given by (\ref{nofw}). In that case 
\begin{equation}
  \label{n2ofw}
  n_{2}^{\min}(\Go)=n_{1}(\Go)+1\le n_{2}(\Go)\le
\max\left\{2,\texttt{round}\left(\frac{T'_{0}\Go}{2\pi}\right)\right\}=n_{2}^{\max}(\Go).
\end{equation}
Thus, for measuring very low frequencies we set $n_{1}(\Go)=1$ and
$n_{2}(\Go)=2$. When $\Go=4\pi/T_{0}$ we have $n_{1}(\Go)=2$, in which case we
must require that $n_{2}(\Go)\ge 3$. It follows that $T'_{0}\ge 3T_{0}/2$,
and therefore, $n_{2}^{\max}(\Go)>n_{1}(\Go)$. In
practice, the experimenter can choose any integer $n_{2}(\Go)$ satisfying
$n_{2}^{\min}(\Go)\le n_{2}(\Go)\le n_{2}^{\max}(\Go)$, once the constant
$T'_{0}\ge 3T_{0}/2$ has been set.

The correction method (\ref{Zcorr}) can be interpreted as a directive to
ignore the transient response over the time interval $[0,T_{1}]$, effectively
taking data during time interval $[T_{1},T_{2}]$. Indeed, using formula
(\ref{impform}), we have
\begin{equation}
  \label{Zexp}
Z_{\rm exp}(\Go)=\frac{2i}{T_{2}(\Go)-T_{1}(\Go)}
\int_{T_{1}(\Go)}^{T_{2}(\Go)}U(t)e^{-i\Go t}dt,\qquad
T_{1,2}(\Go)=\frac{2\pi n_{1,2}(\Go)}{\Go}.
\end{equation}
One convenient choice is $n_{1}=n(\Go)$ and $n_{2}=2n(\Go)$, where $n(\Go)$ is
given by (\ref{nofw}), so that the length of the time interval over which the
data is taken is still $T(\Go)=2\pi n(\Go)/\Go$.
\begin{theorem}
\label{th:Zexpest}
Under assumptions of Theorem~\ref{th:Zest} we have the estimate
\begin{equation}
  \label{Zexpest}
  |Z_{\rm exp}(\Go)-Z(\Go)|\le \frac{T_{1}}{T_{2}-T_{1}}R_{0}\left(1+\frac{\Gth}{T_{1}}\right)e^{-T_{1}/\Gth}+R_{0}e^{-T_{2}/\Gth}.
\end{equation}
\end{theorem}
The proof can be found is in Appendix~\ref{app:proofs}.

To see how formula (\ref{Zexp}) improves the evaluation of the
impedance we take the same Voigt circuit with $R=2k\GO$, $C=200\mu F$, shown in
Figure~\ref{fig:Z0vsZV}, and compare $Z_{\rm exp}(\Go)$ and $Z(\Go)=Z_{V}(\Go)$.
The difference between the Nyquist plots of the two functions
can no longer be visualized as in Figure~\ref{fig:Z0vsZV}, since it is
less that $0.01$\%, according to numerics, when we
choose $n_{1}=n(\Go)$ and $n_{2}=2n(\Go)$.

\section{Non time-invariant elementary Voigt circuits}
\label{sec:EVC}
In this section we examine elementary Voigt circuits made of a resistor and
a capacitor connected in parallel. Our main assumption is that the parameters
$R$ and $C$ of the circuit do not stay constant, but slowly change in time, instead. We
will show that if we apply the impedance measurement recipe from the previous
section to such non time-invariant systems we may obtain impedance curves with
inductive loops.
\subsection{General theory}
Let us start by considering an elementary Voigt circuit consisting of a
resistor $R=R(t)$ and a capacitor $C=C(t)$ connected in parallel. In such a
circuit we have the Ohm's laws for each of the elements
\[
I_{R}(t)=\frac{U(t)}{R(t)},\qquad I_{C}(t)=C(t)\dot{U}(t).
\]
Combining this with the Kirchhoff's law $I(t)=I_{R}(t)+I_{C}(t)$ we obtain the
constitutive relation (i.e. dependence of $I(t)$ on $U(t)$) in the form
\begin{equation}
  \label{elVoigt}
  I(t)=\frac{U(t)}{R(t)}+C(t)\dot{U}(t).
\end{equation}
Since we use the current as
the input we need to solve (\ref{elVoigt}) for $U(t)$:
\begin{equation}
  \label{VoigtIU0}
  U(t)=\int_{-\infty}^{t}\frac{I(s)}{C(s)}\exp\left\{-\int_{s}^{t}
\frac{dx}{\Gth(x)}\right\}ds,\qquad\Gth=RC.
\end{equation}
The quantity $\Gth$ is called the relaxation time of the Voigt circuit.
Formula (\ref{Zexp}) gives the experimentally measured ``impedance'' of a
simple Voigt circuit with non time-invariant elements:
\begin{equation}
  \label{ZIRcor2}
  Z_{\rm exp}(\Go)=\frac{2i}{\GD T}\int_{T_{1}}^{T_{2}}e^{-i\Go t}\int_{0}^{t}\frac{\sin(\Go s)}{C(s)}
  \exp\left\{-\int_{s}^{t}\frac{dx}{\Gth(x)}\right\}dsdt,\quad
  \GD T=T_{2}-T_{1}.
\end{equation}
For future reference we also have the following formula for the impedance of a
non time-invariant capacitor and resistor
\begin{equation}
  \label{Zcap}
Z^{C}_{\rm exp}(\Go)=\frac{2i}{\GD T}\int_{T_{1}}^{T_{2}}e^{-i\Go t}
\int_{0}^{t}\frac{\sin(\Go s)}{C(s)}dsdt=
\frac{2}{\GD T\Go}\int_{T_{1}}^{T_{2}}(e^{-i\Go t}-1)\frac{\sin(\Go t)}{C(t)}dt,
\end{equation}
\begin{equation}
  \label{ZR}
Z^{R}_{\rm exp}(\Go)=\frac{2i}{\GD T}\int_{T_{1}}^{T_{2}}e^{-i\Go t}R(t)\sin(\Go t)dt.
\end{equation}

\subsection{Explicit non time-invariant models}
Let us now examine a particular model of the time dependence of $R(t)$ and
$C(t)$. We
assume that $R$ and $C$ undergo an ``exponential drift'' from $R_{-}$
(resp. $C_{-}$) at $t=-\infty$
to $R_{+}$ (resp. $C_{+}$) at $t=+\infty$:
\begin{equation}
  \label{expdrift}
R(t)=\frac{aR_{-}+R_{+}e^{t/\tau}}{a+e^{t/\tau}},\quad 
C(t)=\frac{bC_{-}+C_{+}e^{t/\tau}}{b+e^{t/\tau}},\quad
a,b>0.  
\end{equation}
The evolution law for $R(t)$ is governed by two parameters: the time scale
$\tau$ and the ``current position'' $a>0$ that tells us how far along
$R(t)$ is on the way from $R_{-}$ to $R_{+}$. It is in 1-1 correspondence with
$R(0)$. The value $a=1$ means $R(0)$ is exactly half-way between $R_{+}$ and
$R_{-}$. Exact same comments apply to $C(t)$. It is easy to calculate explicitly
\begin{equation}
  \label{intheta}
  -\int_{s}^{t}\frac{dx}{\Gth(x)}=\frac{\tau}{\Gth_{+}}\left(\frac{s-t}{\tau}+\Gk_{C}
\ln\left(\frac{\rho_{C}e^{-s/\tau}+1}{\rho_{C}e^{-t/\tau}+1}\right)+
\Gk_{R}\ln\left(\frac{\rho_{R}e^{-s/\tau}+1}{\rho_{R}e^{-t/\tau}+1}
\right)\right), 
\end{equation}
where
\[
\Gk_{C}=\frac{(a-\rho_{C})(b-\rho_{C})}{\rho_{C}(\rho_{C}-\rho_{R})},\quad
\Gk_{R}=\frac{(a-\rho_{R})(b-\rho_{R})}{\rho_{R}(\rho_{R}-\rho_{C})},\quad
\rho_{R}=\frac{aR_{-}}{R_{+}},\quad\rho_{C}=\frac{bC_{-}}{C_{+}},\quad\rho_{R}\not=\rho_{C}.
\]
We note a relation
\begin{equation}
  \label{ABrel}
  \Gk_{C}+\Gk_{R}=1-\frac{\Gth_{+}}{\Gth_{-}},\qquad\Gth_{\pm}=C_{\pm}R_{\pm}.
\end{equation}

If $\rho_{R}=\rho_{C}=\rho$, then
\begin{equation}
  \label{intheta0}
  -\int_{s}^{t}\frac{dx}{\Gth(x)}=\frac{\tau}{\Gth_{+}}\left(\frac{s-t}{\tau}+
\Ga\ln\left(\frac{\rho e^{-s/\tau}+1}{\rho e^{-t/\tau}+1}\right)
+\frac{\Gb(e^{-s/\tau}-e^{-t/\tau})}{(\rho e^{-s/\tau}+1)(\rho e^{-t/\tau}+1)}\right), 
\end{equation}
where
\[
\Ga=1-\frac{\Gth_{+}}{\Gth_{-}},\qquad\Gb=\frac{(a-\rho)(b-\rho)}{\rho}.
\]

A different drift model
\begin{equation}
  \label{model2}
  R(t)=R_{0}e^{-t/\tau}+R_{\infty}(1-e^{-t/\tau}),\qquad C(t)=C_{0}e^{-t/\tau}+C_{\infty}(1-e^{-t/\tau})
\end{equation}
is a limiting case of our model (\ref{expdrift}) with $R_{+}=R_{\infty}$, $C_{+}=C_{\infty}$, and 
\[
  a,b\to 0,\quad R_{-},C_{-}\to\infty,\quad aR_{-}\to
  R_{0}-R_{\infty},\quad bC_{-}\to C_{0}-C_{\infty}.
\]
In this limit, however, the restriction that $R_{-}$ and $C_{-}$ are positive
are no longer required, since $R(t)$ and $C(t)$ are required to be positive
only for $t\ge 0$. In this drift model we can still use (\ref{intheta}) with
\[
\rho_{R}\to\frac{R_{0}}{R_{\infty}}-1,\quad\rho_{C}\to\frac{C_{0}}{C_{\infty}}-1,\quad
\Gk_{R}\to\frac{\rho_{R}}{\rho_{R}-\rho_{C}},\quad\Gk_{C}\to\frac{\rho_{C}}{\rho_{C}-\rho_{R}},
\]
provided $\rho_{R}\not=\rho_{C}$. If $\rho_{R}=\rho_{C}=\rho$, then we use
formula (\ref{intheta0}) with $\Ga=1$ and $\Gb=\rho$.

Once, $\int_{s}^{t}\Gth(x)^{-1}dx$ has been evaluated, we compute
\begin{equation}
\label{EVCED}
U(t)=\frac{I_{0}}{C_{+}}\int_{0}^{t}\sin(\Go s)\frac{1+be^{-s/\tau}}{\rho_{C}e^{-s/\tau}+1}
\exp\left\{-\int_{s}^{t}\frac{dx}{\Gth(x)}\right\}ds.
\end{equation}
The experimenatally measured impedance $Z_{\exp}(\Go)$ is then
computed by means of formula (\ref{Zexp}). The numerical evaluation of
$Z_{\exp}(\Go)$ presents several challenges, since 
parameter $\tau$ must be very large, relaxation
times $\Gth_{\pm}$ are typically very small and $\Go$ can range from
$10^{-6}$Hz to $10^{6}$Hz, making some integrands highly
oscillatory. These issues are addressed in Appendix~\ref{app:num}.
\begin{figure}[t]
  \centering
  \includegraphics[scale=0.3]{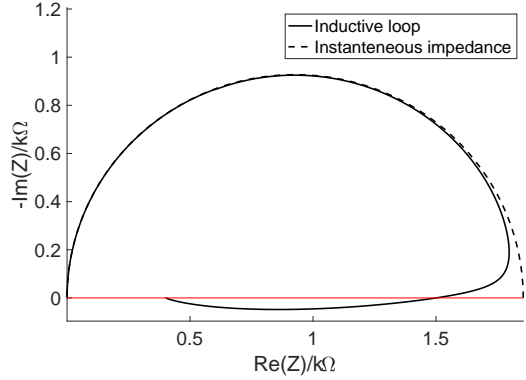}
  \caption{An inductive loop in a non time-invariant Voigt element.}
  \label{fig:VgtdrftRC}
\end{figure}
Figure~\ref{fig:VgtdrftRC} shows an inductive loop in the Nyquist impedance
plot of a non time-invariant Voigt element, with $R(t)$ and $C(t)$ given
by (\ref{expdrift}), where $R_{+}=0.4k\GO$, $R_{-}=2k\GO$, $a=10$,
$C_{+}=1.6mF$, $C_{-}=0.8mF$, $b=6$, $\tau=600s$. The figure shows that at high
frequencies the impedance is very well captured by the impedance of the
time-invariant Voigt element with $R=R(0)$, and $C=C(0)$.

\section{Slow parameter drift asymptotics}
\label{sec:asym}
Assume that $1/\tau=\Ge$ is a small parameter and that $\Go=\Ge\GO$, where the
rescaled frequency $\GO$ is assumed to be fixed. We need to find the
asymptotics of $Z(\Ge\GO)$ as $\Ge\to 0$ and see if $\Im(Z(\Ge\GO))$ can
indeed be positive.  We now assume that $R(t)=R_{0}(\Ge t)$ and
$C(t)=C_{0}(\Ge t)$, where $R_{0}(x)$ and $C_{0}(x)$ are strictly positive,
bounded smooth functions with bounded first derivatives on $\bb{R}$. Our goal
is not only to obtain the asymptotics of $Z(\Go)$, defined by (\ref{impform}),
as $\Ge\to 0$, but estimate the error between the actual impedance and its
asymptotic approximation. We obtain for the elementary Voigt circuit, using
(\ref{VoigtIU0})
\[
Z_{\Ge}(\Go)=\frac{2i}{\GD T}\int_{T_{1}}^{T_{2}}e^{-i\Go t}\int_{0}^{t}\frac{\sin(\Go s)}{C_{0}(\Ge s)}
\exp\left\{-\int_{s}^{t}\frac{d\tau}{\Gth(\Ge\tau)}\right\}dsdt,\quad \GD T=T_{2}-T_{1}.
\]
When the drift time scale $\tau=1/\Ge$ is large, the formula for
$Z_{\Ge}(\Go)$ above can be simplified:
\begin{equation}
  \label{twoscale}
  Z_{\Ge}(\Go)=\frac{i\GO}{\pi\GD n(\Go)}
  \int_{2\pi n_{1}(\Go)/\GO}^{2\pi n_{2}(\Go)/\GO}e^{-i\GO\eta}
\im\left(\frac{R_{0}(\eta)e^{i\GO\eta}}{1+i\Go\Gth_{0}(\eta)}\right)d\eta
+ O(\Ge),
\end{equation}
where the order $\Ge$ error $O(\Ge)$ is \emph{uniform over the entire
  frequency spectrum}. 

 Formula
(\ref{twoscale}) shows that if the functions $\Gth_{0}(\eta)$ and
$R_{0}(\eta)$ are constants, then the approximation becomes exact, as it
reduces to the classical impedance of the elementary Voigt circuit.
One more simplification is possible in the asymptotics $\Ge\to
0$. Specifically, we can replace $\Gth_{0}(\eta)$ by $\Gth(0)$ in
(\ref{twoscale}). Our final asymptotics is
\begin{equation}
  \label{twoscale0}
  Z^{*}_{\Ge}(\Go)=Z(\Go,\GO)=\frac{i\GO}{\pi\GD n(\Go)}
  \int_{2\pi n_{1}(\Go)/\GO}^{2\pi n_{2}(\Go)/\GO}e^{-i\GO\eta}
\im\left(\frac{R_{0}(\eta)e^{i\GO\eta}}{1+i\Go\Gth_{0}(0)}\right)d\eta.
\end{equation}
\begin{figure}[t]
  \centering
  \includegraphics[scale=0.25]{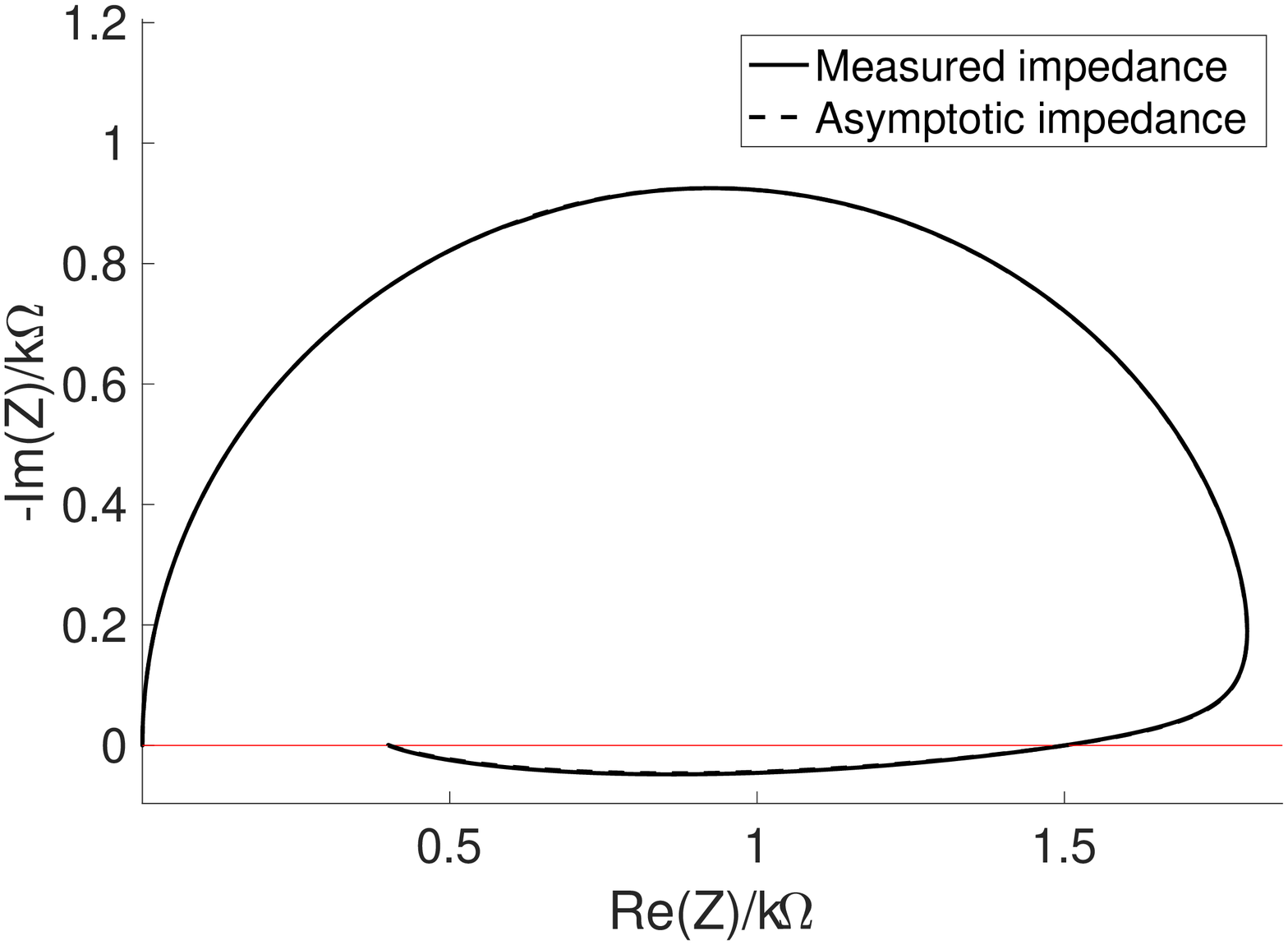}~~~~~~~~~~~
\includegraphics[scale=0.25]{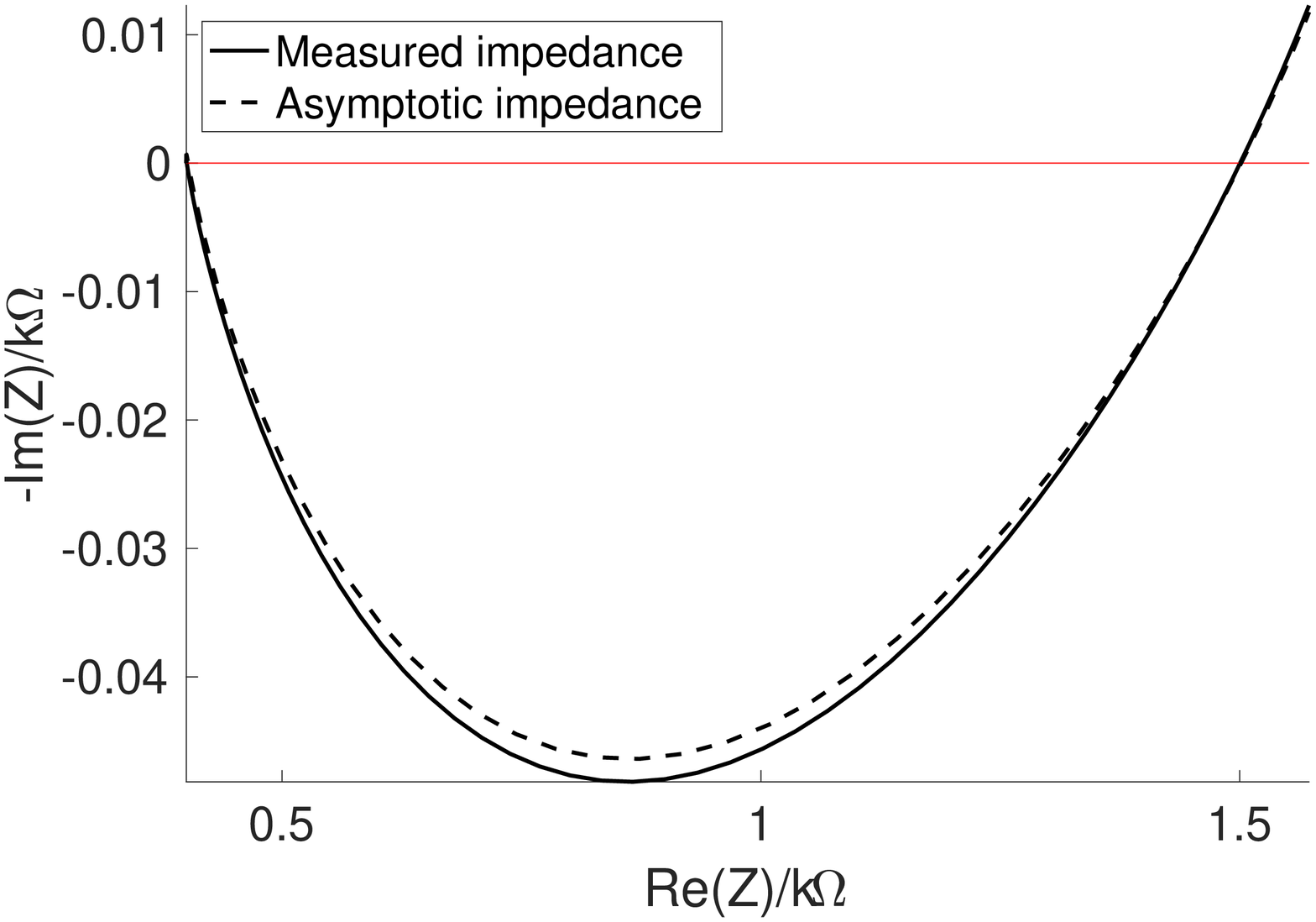}
  \caption{The quality of the asymptotics of an inductive loop.}
  \label{fig:asymp}
\end{figure}

For numerical evaluation of (\ref{twoscale0}) we choose
$n_{1}=n(\Go)$, $n_{2}=2n(\Go)$ and change variables of integration
$\eta=sn/\GO$. Then, using the representation $\im(z)=(z-\bra{z})/2i$
we obtain the formula we use in the Matlab code to make Fig.~\ref{fig:asymp}:
\begin{equation}
  \label{indloopeffect}
  Z_{\eps}^{*}(\Go)=\nth{2\pi}\int_{2\pi}^{4\pi}
\frac{R_{0}\left(\frac{sn}{\GO}\right)}{1+i\Go\Gth_{0}(0)}ds-\nth{2\pi}\int_{2\pi}^{4\pi}
\frac{R_{0}\left(\frac{sn}{\GO}\right)e^{-2ins}}{1-i\Go\Gth_{0}(0)}ds.
\end{equation}
For the model (\ref{expdrift}) we compute
\[
\nth{2\pi}\int_{2\pi}^{4\pi}R_{0}\left(\frac{sn}{\GO}\right)ds=
R_{+}+\frac{(R_{+}-R_{-})\GO}{2\pi n}\ln\left(1+\frac{a(e^{-\frac{4\pi n}{\GO}}
-e^{-\frac{2\pi n}{\GO}})}{1+ae^{-\frac{2\pi n}{\GO}}}\right).
\]
In the high frequency regime
\[
R_{0}\left(\frac{sn}{\GO}\right)=R_{0}\left(\frac{\Ge s T_{0}}{2\pi}\right)
\]
We can therefore, use a linear approximation $R_{0}(x)\approx b_{R}+m_{R}x$, when
$x\in[\Ge T_{0},2\Ge T_{0}]$, where the parameters $b_{R}$ and $m_{R}$ are
found from the least squares fit. In that case
\[
\nth{2\pi}\int_{2\pi}^{4\pi}R_{0}\left(\frac{sn}{\GO}\right)e^{-2ins}ds\approx
\frac{im_{R}}{2\GO}.
\]
Figure~\ref{fig:asymp} shows the Nyquist plot of $Z_{\Ge}^{*}(\Go)$ superimposed on
the computed ``measured'' impedance corresponding to the model
(\ref{expdrift}). The difference between the two graphs is not
detectable at the scale of the entire Nyquist plot, shown in the left panel of
the figure. The right panel shows the blown-up portion of the plot containing
the inductive loop. The overall relative error of the asymptotic approximation
(\ref{twoscale0}) is about 0.1\%. In the figure we used (\ref{expdrift}) with
the same parameter values as in Figure~\ref{fig:VgtdrftRC}.

Now, if we have a general electrochemical system exhibiting
parameter drift, then the measured impedance can be represented
as
\begin{equation}
  \label{Zcorrasym}
Z_{\rm exp}(\Go)=\frac{i\GO}{\pi( n_{2}(\Go)-n_{1}(\Go))}
\int_{2\pi n_{1}(\Go)/\GO}^{2\pi n_{2}(\Go)/\GO}e^{-i\GO\eta}
\im\left(Z(\Go,\eta)e^{i\GO\eta}\right)d\eta +O(\Ge),
\end{equation}
where for each $\eta\in\bb{R}$ the function $\Go\mapsto Z(\Go,\eta)$ is a
classical (instantaneous) impedance function, provided there is no
leading capacitance in the ECM. In the presence of the leading
capacitance, the term
\[
Z_{\rm exp}^{C}(\Go)=\frac{i\GO}{\pi (n_{2}(\Go)-n_{2}(\Go))}\int_{2\pi n_{1}(\Go)/\GO}^{2\pi n_{2}(\Go)/\GO}
(e^{-i\GO\eta}-1)\frac{\sin(\GO\eta)}{i\Go C_{0}(\eta)}d\eta
\]
must be added to the \rhs\ of (\ref{Zcorrasym}).

If we approximate the instantaneous impedance function $Z(\Go,\Ge t)$ by a
finite Voigt circuit ECM \cite{aogr92}
\[
Z(\Go,\eta)=R_{0}(\eta)+\sum_{j=1}^{N}\frac{R_{j}(\eta)}{1+i\Go\Gth(\eta)},\]
then, up to a uniform, over the entire spectrum, order $\Ge$ error, we
have
\begin{equation}
  \label{Zasym0}
    Z_{\rm exp}(\Go)=\frac{i\GO}{\pi( n_{2}(\Go)-n_{1}(\Go))}
\int_{2\pi n_{1}(\Go)/\GO}^{2\pi n_{2}(\Go)/\GO}e^{-i\GO\eta}
\im\left(Z^{0}(\Go,\eta)e^{i\GO\eta}\right)d\eta +O(\Ge),
\end{equation}
where
\begin{equation}
  \label{Z0ECM}
  Z^{0}(\Go,\eta)=R_{0}(\eta)+\sum_{j=1}^{N}\frac{R_{j}(\eta)}{1+i\Go\Gth(0)}.
\end{equation}
If $R_{j}(\eta)$ in (\ref{Z0ECM}) are modeled according to (\ref{model2}),
\begin{equation}
  \label{Rexpdrft}  
  R_{j}(\eta)=R_{j}(0)+\GD R_{j}\left(1-e^{-\mu_{j}\eta}\right),\quad j=0,\ldots,N,
\end{equation}
then the integral in (\ref{Zasym0}) can be computed explicitly:
\begin{equation}
  \label{Zasymexpl}
  Z_{\rm exp}(\Go)=\sum_{j=0}^{N}\left\{\frac{R_{j}(0)+\GD R_{j}}{1+i\Go\Gth_{j}(0)}
+\GD R_{j}F\left(\frac{\Ge\mu_{j}}{\Go},\Go\Gth_{j}(0)\right)\right\}+O(\Ge),
\end{equation}
where $\Gth_{0}=0$, and
\[
F(W,v)=i\frac{e^{-2\pi n_{1}W}-e^{-2\pi n_{2}W}}{\pi(n_{2}-n_{1})W}
\frac{v(W+i)-1}{(1+v^{2})(W+2i)}.
\]
Typically, one would choose $n_{1}(\Go)=n(\Go)$ and $n_{2}(\Go)=2n(\Go)$,
where $n(\Go)$ is given by (\ref{nofw}). At low frequencies $n(\Go)=1$, while
at high frequencies $2\pi n(\Go)/\Go$ can be replaced with a constant $T_{0}$.

\section{Conclusions and discussion}
In this paper we have proposed an explanation of the inductive loop phenomena
observed in experiments. The hypothesis is that the act of the impedance
measurement of an electrochemical system might speed up slow processes, such
as corrosion and charge diffusion, altering the properties of the system on
time scales commensurate with the inverses of the frequencies at which the
impedance is measured. The modeling and analysis of these processes can lead
to specific drift laws that can be used to validate our general theory.

A natural question is whether our theory makes it apparent how the time-dependent
nature of the system causes the occurrence of inductive loops. By way of the
answer we refer to formula (\ref{indloopeffect}) and set $\Go=0$,
$n(\Go)=1$. Then we obtain
\begin{equation}
  \label{indlpimZ}
\im(Z(\Ge\GO))\approx\nth{2\pi}\int_{2\pi}^{4\pi}
R_{0}\left(\frac{s}{\GO}\right)\sin(2s)ds
\end{equation}
While $R_{0}(u)$ is a strictly positive, smooth real function, the sign of the
integral in (\ref{indlpimZ}) is the outcome of balancing positive and negative
contributions of the oscillatory integrand. For example, if $R_{0}(u)$ is a
decreasing function, then positive contributions will always be larger than the
negative ones and inductive loops will be present. Conversely, if $R_{0}(u)$ is an
increasing function, no inductive loops will be produced. If the slowly
oscillating sinusoidal current input causes a corresponding in-phase oscillation
of $R_{0}$, the inductive loop effect could be several times stronger.

One open question, not directly addressed in the paper is capturing the
time scale $\tau$ over which the system's properties change. A crude estimate
would be the inverse frequency at which the imaginary part of the impedance
hits zero. For example, in the simulation corresponding to
Figure~\ref{fig:VgtdrftRC} the time scale $\tau=600s$, while
$2\pi/\Go_{0}=610s$, where $\im(Z(\Go_{0}))=0$.
Our simulation of the time-dependent elementary Voigt circuit shows that the
deviations from the classical model start at frequencies $\Go$ as large as two
orders of magnitude over $\tau$. They become vividly pronounced at frequencies
on the order of $1/\tau$ and persist over frequencies an order of magnitude
lower. 

Finally, whether or not the inductive loop data can give additional
information about the system beyond the time scale $\tau$ depends on whether
the drift model (\ref{Rexpdrft}) is acceptable. In this case one should be
able to use the algorithm in \cite{grab_Stielt} applied to the high frequency
data to compute parameters $R_{j}(0)$, $\Gth_{j}(0)$ and $N$. The intermediate
and low frequency data can then be used in a non-linear least squares fit to to
estimate drift parameters $\mu_{j}$ and $\GD R_{j}$. Future research into the
inductive loops should address these questions.

\medskip

\textbf{Acknowledgments.} This material is based upon work supported by the National
Science Foundation under Grant No. DMS-2005538.

\appendix
\section{Mathematical proofs}
\label{app:proofs}
\subsection{Proof of Theorem~\ref{th:Zest} and \ref{th:Zexpest}}
For $I(t)=\chi_{[0,T(\Go)]}(t)\sin(\Go t)$ we have,
making a change of variables $s=t-\tau$,
\[
U(t)=\rho_{0}\sin(\Go t)+\int_{0}^{t}K(s)\sin(\Go(t-s))ds,\qquad 0\le t\le T(\Go).
\]
We can then write
\[
U(t)=\rho_{0}\sin(\Go t)+\int_{0}^{\infty}\chi_{(s,+\infty)}(t)K(s)\sin(\Go(t-s))ds,
\]
so that we can substitute this into (\ref{impform}) and switch the order of
integration. We obtain
\[
Z_{0}(\Go)=\rho_{0}+\frac{2i}{T(\Go)}\int_{0}^{T(\Go)}K(s)
\left(\int_{s}^{T(\Go)}\sin(\Go(t-s))e^{-i\Go t}dt\right)ds.
\]
Computing the inner integral we obtain
\[
Z_{0}(\Go)=\rho_{0}+\int_{0}^{T(\Go)}K(s)\left(e^{-i\Go s}
-\frac{\Go se^{-i\Go s}-\sin(\Go s)}{2\pi n}\right)ds.
\]
Using formula (\ref{trueZ}) for the true impedance, we compute
\[
Z(\Go)-\frac{1}{2\pi n(\Go)}\im(Z(\Go))-\frac{i\Go}{2\pi n(\Go)}Z'(\Go)=\rho_{0}+
\int_{0}^{\infty}K(s)\left(e^{-i\Go s}+\frac{\sin(\Go s)-\Go se^{-i\Go s}}{2\pi n}\right)ds.
\]
Therefore
\begin{equation}
  \label{remainder}
Z(\Go)-\frac{\im(Z(\Go))}{\Go T(\Go)}-\frac{i}{T(\Go)}Z'(\Go)-Z_{0}(\Go)=\int_{T(\Go)}^{\infty}
K(s)\left(e^{-i\Go s}+\frac{\sin(\Go s)-\Go se^{-i\Go s}}{2\pi n}\right)ds.
\end{equation}
For $s\ge T(\Go)=2\pi n/\Go$ we have $\Go s\ge 2\pi n$. Denoting $x=\Go s$ and
$A=1/(2\pi n)$, we estimate
\[
|e^{-ix}+A(\sin x-xe^{-ix})|\le|1-Ax|+A=Ax-1+A\le Ax,
\]
since $Ax\ge 1$ and $A\le 1/(2\pi)<1$. Hence,
\[
\left|Z_{0}(\Go)-Z(\Go)+\frac{\im(Z(\Go))+i\Go Z'(\Go)}{2\pi n(\Go)}\right|
\le\nth{T(\Go)}\int_{T(\Go)}^{\infty}s|K(s)|ds.
\]
Using the exponential decay (\ref{decay}) of the memory kernel we obtain the estimate
\begin{equation}
  \label{Zest}
  \left|Z_{0}(\Go)-Z(\Go)+\frac{\im(Z(\Go))+i\Go Z'(\Go)}{2\pi n(\Go)}\right|
\le R_{0}\left(1+\frac{\Gth}{T(\Go)}\right)e^{-\frac{T(\Go)}{\Gth}}, 
\end{equation}
proving (\ref{correst}). 

Let us now prove Theorem~\ref{th:Zexpest}. We compute,
\[
Z_{\rm exp}(\Go)-Z(\Go)=
\frac{T_{1}}{T_{2}-T_{1}}\int_{T_{1}}^{T_{2}}
K(s)\left(e^{-i\Go s}+\frac{\sin(\Go s)-\Go se^{-i\Go s}}{2\pi n_{1}}\right)ds
-\int_{T_{2}}^{\infty}K(s)e^{-i\Go s}ds.
\]
Uisng estimate (\ref{Zest}) we obtain
\[
|Z_{\rm exp}(\Go)-Z(\Go)|\le\frac{T_{1}(\Go)}{T_{2}(\Go)-T_{1}(\Go)}R_{0}\left(1+\frac{\Gth}{T_{1}(\Go)}\right)
e^{-\frac{T_{1}(\Go)}{\Gth}}+R_{0}e^{-\frac{T_{2}(\Go)}{\Gth}}.
\]

\subsection{Proof of the asymptotic foormula (\ref{Zcorrasym})}

Let us analyze the asymptotic behavior of $Z_{\Ge}(\Go)$. We first change
variables in the innermost integral $\Gs=\Ge\tau$:
\[
Z_{\Ge}(\Go)=\frac{2i}{\GD T}\int_{T_{1}}^{T_{2}}e^{-i\Go t}\int_{0}^{t}\frac{\sin(\Go s)}{C_{0}(\Ge s)}
\exp\left\{-\nth{\Ge}\int_{\Ge s}^{\Ge t}\frac{d\Gs}{\Gth(\Gs)}\right\}dsdt.
\]
Next we change variables $\xi=\Ge s$ in the integral with respect to the $s$ variable:
\[
Z_{\Ge}(\Go)=\frac{2i}{\GD T\Ge}\int_{T_{1}}^{T_{2}}e^{-i\Go t}\int_{0}^{t\Ge}\frac{\sin(\Go\xi/\Ge)}
{C_{0}(\xi)}\exp\left\{-\nth{\Ge}\int_{\xi}^{\Ge t}\frac{d\Gs}{\Gth(\Gs)}\right\}d\xi dt.
\]
Finally, we change variables in the outermost integral $\eta=\Ge t$:
\[
Z_{\Ge}(\Go)=\frac{2i}{\GD T\Ge^{2}}\int_{T_{1}\Ge}^{T_{2}\Ge}e^{-i\GO\eta}\int_{0}^{\eta}\frac{\sin(\GO\xi)}
{C_{0}(\xi)}\exp\left\{-\nth{\Ge}\int_{\xi}^{\eta}\frac{d\Gs}{\Gth(\Gs)}\right\}d\xi d\eta,
\]
where $\GO=\Go/\Ge$ could be large, when $\Go$ is not very small, but
could also be of order 1, when $\Go$ is of order $\Ge$. 

The main approximation idea is to realize that
\[
E_{\Ge}(\xi,\eta)=\exp\left\{-\nth{\Ge}\int_{\xi}^{\eta}\frac{d\Gs}{\Gth(\Gs)}\right\}
\]
is exponentially small when $\eta-\xi$ is not very small. However, when
$\eta\approx\xi$, then $\Gth(\Gs)\approx\Gth(\eta)$, which means that
\[
E_{\Ge}(\xi,\eta)\approx E^{0}_{\Ge}(\xi,\eta)=\exp\left\{-\frac{\eta-\xi}{\Gth(\eta)\Ge}\right\}.
\]
To make this approximation quantitative we use the inequality
\[
\frac{e^{x_{2}}-e^{x_{1}}}{x_{2}-x_{1}}< e^{\max\{x_{1},x_{2}\}},
\]
which is a consequence of convexity of the exponential function. Hence,
\[
|E_{\Ge}(\xi,\eta)-E^{0}_{\Ge}(\xi,\eta)|\le
\frac{L_{\Gth^{-1}}(\eta-\xi)^{2}}{2\Ge}\exp\left\{-\frac{\eta-\xi}{M_{\Gth}\Ge}\right\}=
\Ge LB\left(\frac{\eta-\xi}{\Ge}\right),
\]
where 
\[
M_{\Gth}=\max_{t\in\bb{R}}\Gth(t),\qquad 
L_{\Gth^{-1}}=\max_{t\in\bb{R}}\left|\left(\nth{\Gth(t)}\right)'\right|,\qquad
B(x)=\frac{x^{2}}{2}\exp\left\{-\frac{x}{M_{\Gth}}\right\}.
\]
Thus, we can replace $Z_{\Ge}(\Go)$ with
its approximation
\[
Z_{\Ge}^{(1)}(\Go)=\frac{2i}{\GD T\Ge^{2}}\int_{T_{1}\Ge}^{T_{2}\Ge}e^{-i\GO\eta}\int_{0}^{\eta}\frac{\sin(\GO\xi)}
{C_{0}(\xi)}E_{\Ge}^{0}(\xi,\eta)d\xi d\eta.
\]
Moreover, we also have
\[
|Z_{\Ge}(\Go)-Z_{\Ge}^{(1)}(\Go)|\le\frac{2L_{\Gth^{-1}}}{m_{C}\GD T\Ge}\int_{T_{1}\Ge}^{T_{2}\Ge}\int_{0}^{\eta}
B\left(\frac{\eta-\xi}{\Ge}\right)d\xi d\eta,
\]
where
\[
m_{C}=\min_{t\in\bb{R}}C(t).
\]
Changing variables in the inner integral $x=(\eta-\xi)/\Ge$ we obtain
\[
|Z_{\Ge}(\Go)-Z_{\Ge}^{(1)}(\Go)|\le\frac{2L_{\Gth^{-1}}\Ge}{m_{C}}\int_{0}^{\infty}B(x)dx=
\frac{\Ge M_{\Gth}^{3}L_{\Gth^{-1}}}{m_{C}}.
\]
The function $E^{0}_{\Ge}(\xi,\eta)$ is exponentially small when $\xi$ is not
very close to $\eta$. However, when $\xi$ is very close to $\eta$ we can
replace $C_{0}(\xi)$ with $C_{0}(\eta)$. Hence we have
\[
Z_{\Ge}^{(1)}(\Go)\approx Z_{\Ge}^{(2)}(\Go)=\frac{2i}{\GD T\Ge^{2}}\int_{T_{1}\Ge}^{T_{2}\Ge}
\frac{e^{-i\GO\eta}}{C_{0}(\eta)}\int_{0}^{\eta}\sin(\GO\xi)E_{\Ge}^{0}(\xi,\eta)d\xi d\eta.
\]
Moreover,
\[
|Z_{\Ge}^{(1)}(\Go)-Z_{\Ge}^{(2)}(\Go)|\le\frac{2L_{C^{-1}}}{\GD T\Ge^{2}}\int_{T_{1}\Ge}^{T_{2}\Ge}
\int_{0}^{\eta}(\eta-\xi)\exp\left\{-\frac{\eta-\xi}{\Ge M_{\Gth}}\right\}d\xi d\eta,
\]
where
\[
L_{C^{-1}}=\max_{t\in\bb{R}}\left|\left(\nth{C(t)}\right)'\right|.
\]
Changing variables $x=(\eta-\xi)/\Ge$ in the inner integral we obtain the
bound
\[
|Z_{\Ge}^{(1)}(\Go)-Z_{\Ge}^{(2)}(\Go)|\le\frac{2L_{C^{-1}}}{\GD T}\int_{T_{1}\Ge}^{T_{2}\Ge}
\int_{0}^{\infty}x\exp\left\{-\frac{x}{M_{\Gth}}\right\}dxd\eta=2L_{C^{-1}}M_{\Gth}^{2}\Ge.
\]

Now, the inner integral in $Z_{\Ge}^{(2)}(\Go)$ can be evaluated explicitly:
\[
\int_{0}^{\eta}\sin(\GO\xi)\exp\left\{-\frac{\eta-\xi}{\Gth(\eta)\Ge}\right\}d\xi=
\frac{\Ge\Gth\sin(\GO\eta)-\Ge^{2}\Gth^{2}\GO\cos(\GO\eta)+\Ge^{2}\Gth^{2}\GO e^{-\eta/(\Ge\Gth)}}
{1+\Ge^{2}\Gth^{2}\GO^{2}}.
\]
We observe that the first two terms can be combined nicely, and
\[
Z_{\Ge}^{(2)}(\Go)=\frac{2i}{\GD T\Ge}\int_{T_{1}\Ge}^{T_{2}\Ge}
e^{-i\GO\eta}\im\left(\frac{R_{0}(\eta)e^{i\GO\eta}}{1+i\Go\Gth_{0}(\eta)}\right)d\eta
+\GD_{\Ge}(\Go)=Z_{\Ge}^{(3)}(\Go)+\GD_{\Ge}(\Go),
\]
where
\[
\GD_{\Ge}(\Go)=\frac{2i\GO}{\GD T}\int_{T_{1}\Ge}^{T_{2}\Ge}e^{-i\GO\eta}
\frac{\Gth^{2}e^{-\eta/(\Ge\Gth)}}{C_{0}(\eta)(1+\Ge^{2}\Gth^{2}\GO^{2})}d\eta.
\]
We estimate
\[
  |\GD_{\Ge}(\Go)|\le
  2\GO\Ge\max_{\eta\in[T_{1}\Ge,T_{2}\Ge]}\frac{\Gth(\eta)^{2}e^{-\eta/\Ge\Gth(\eta)}}
  {C_{0}(\eta)(1+\Ge^{2}\Gth(\eta)^{2}\GO^{2})}\le
  M_{R}e^{-T_{1}/M_{\Gth}}.\] $Z_{\Ge}^{(3)}(\Go)$ can be written as a
two-scale impedance function (\ref{twoscale}).

In order to prove a simplified formula (\ref{twoscale0})
we estimate
\[
|Z^{(3)}_{\Ge}(\Go)-Z^{*}_{\Ge}(\Go)|\le\frac{\GO}{\pi\GD n(\Go)}
\frac{M_{R}L_{\Gth}}{1+m_{\Gth}^{2}\Go^{2}}
\int_{2\pi n_{1}(\Go)/\GO}^{2\pi n_{2}(\Go)/\GO}\Go\eta d\eta=
\frac{2\pi\Ge(n_{1}+n_{2})M_{R}L_{\Gth}}{1+m_{\Gth}^{2}\Go^{2}}.
\]
If we choose $n_{1}=n$ and $n_{2}=2n$, then in the regime $n(\Go)=1$ we get
the bound
\[
|Z^{(3)}_{\Ge}(\Go)-Z^{*}_{\Ge}(\Go)|\le 6\pi\Ge M_{R}L_{\Gth}.
\]
If $n(\Go)>1$, then $n(\Go)=T_{0}\Go/2\pi$, and we obtain
\[
|Z^{(3)}_{\Ge}(\Go)-Z^{*}_{\Ge}(\Go)|\le\frac{3\Ge T_{0}\Go M_{R}L_{\Gth}}{1+m_{\Gth}^{2}\Go^{2}}\le
\frac{3\Ge T_{0}M_{R}L_{\Gth}}{2m_{\Gth}}.
\]
Hence, we obtain a uniform in the entire spectrum bound
\[
|Z_{\Ge}(\Go)-Z^{*}_{\Ge}(\Go)|\le M\Ge
\]
for some constant $M$ that depends on the parameter functions $R_{0}(s)$ and
$C_{0}(s)$, and that scales like $R\Gth$.

\section{Numerical evaluation of the impedance in explicit drift  models}
\label{app:num}
The key to the effective numerical computation of $Z_{\exp}(\Go)$ for
explicit drift models (\ref{expdrift}) and (\ref{model2}) is a
non-dimensionalization of all quantities. We therefore begin by rescaling the
variable of integration in the definition (\ref{ZIRcor2}) of $Z(\Go)$, $t=n\hat{t}/\Go$.
\[
Z(\Go)=\frac{i}{I_{0}\pi}\int_{2\pi}^{4\pi}U\left(\frac{n\hat{t}}{\Go}\right)e^{-in\hat{t}}d\hat{t},
\]
where we have chosen $n_{1}=n(\Go)$, $n_{2}=2n(\Go)$, and where
$n(\Go)$ is given by (\ref{nofw}). We then observe that $U(t)$ has the form
\begin{equation}
  \label{Udrift}
 U(t)=\frac{I_{0}}{C_{+}}\int_{0}^{t}\sin(\Go s)f\left(\frac{s}{\tau}\right)
e^{\frac{\tau}{\Gth_{+}}\left[g\left(\frac{s}{\tau}\right)-g\left(\frac{t}{\tau}\right)\right]}ds,
\end{equation}
where
\[
  f(u)=\frac{be^{-u}+1}{\rho_{C}e^{-u}+1},\qquad
  g(u)=u+\Gk_{R}\ln(\rho_{R}e^{-u}+1)+\Gk_{C}\ln(\rho_{C}e^{-u}+1),
\]
provided $\rho_{C}\not=\rho_{R}$. If $\rho_{C}=\rho_{R}=\rho$, then we use
\[
  g(u)=u+\Ga\ln(\rho e^{-u}+1)-\frac{\Gb}{\rho(\rho e^{-u}+1)}.
\]
We therefore, rescale the variable of integration $s=n\hat{s}/\Go$ in (\ref{Udrift}):
\[
U\left(\frac{n\hat{t}}{\Go}\right)=\frac{I_{0}n}{C_{+}\Go}\int_{0}^{\hat{t}}\sin(n\hat{s})
f\left(\frac{n\hat{s}}{\Go\tau}\right)
e^{\frac{\tau}{\Gth_{+}}\left[g\left(\frac{n\hat{s}}{\Go\tau}\right)-g\left(\frac{n\hat{t}}{\Go\tau}\right)\right]}d\hat{s}
\]
Hence, we obtain the formula we use in our Matlab code
\begin{equation}
  \label{Zcomp}
  Z(\Go)=\frac{in}{\pi C_{+}\Go}\int_{2\pi}^{4\pi}e^{-in\hat{t}}\left(\int_{0}^{\hat{t}}\sin(n\hat{s})
f\left(\frac{n\hat{s}}{\Go\tau}\right)
e^{\frac{\tau}{\Gth_{+}}\left[g\left(\frac{n\hat{s}}{\Go\tau}\right)-g\left(\frac{n\hat{t}}{\Go\tau}\right)\right]}d\hat{s}\right)d\hat{t}.
\end{equation}
In the frequency band where $n(\Go)/\Go\approx T_{0}$, we approximate, assuming that $\tau$ is
large,
\[
f\left(\frac{n\hat{s}}{\Go\tau}\right)\approx b_{f}+m_{f}\frac{n\hat{s}}{\Go\tau},\qquad
g\left(\frac{n\hat{s}}{\Go\tau}\right)\approx b_{g}+
m_{g}\frac{n\hat{s}}{\Go\tau},
\]
where the slopes $m_{f}$ and $m_{g}$, and intercepts $b_{f}$ and $b_{g}$ are
found from linear least squares fits of $f(x)$ and $g(x)$, $x\in[T_{0}/\tau,2T_{0}/\tau]$.
Then, all integrals in (\ref{Zcomp}) can be computed explicitly. Up to
exponentially small terms 
\[
Z(\Go)\approx \frac{b_{f}R_{+}}{m_{g}+i\Go\Gth_{+}}+\nth{2\Go\tau}
\frac{m_{f}R_{+}}{m_{g}+i\Go\Gth_{+}}\left(3T_{0}\Go+3i-\frac{4im_{g}^{2}}{m_{g}^{2}+\Go^{2}\Gth_{+}^{2}}\right),\quad
n(\Go)>1,
\]
where in the final expression we replaced $n(\Go)$ with $T_{0}\Go/(2\pi)$,
which is valid exactly in the high frequency regime.

\def\cprime{$'$} \ifx \cedla \undefined \let \cedla = \c \fi\ifx \cyr
  \undefined \let \cyr = \relax \fi\ifx \cprime \undefined \def \cprime
  {$\mathsurround=0pt '$}\fi\ifx \prime \undefined \def \prime {'}
  \fi\def\Ya{Ya}

\end{document}